# Circularly polarized light detector based on ferromagnet/semiconductor junctions

H. Ikeda, N. Nishizawa, K. Nishibayashi, and H. Munekata

Imaging Science and Engineering Laboratory, Tokyo Institute of Technology,
4259-J3-15 Nagatsuta-cho, Midori-ku, Yokohama 226-8503, Japan

Helicity-dependent photocurrent $\Delta I$ has been detected successfully under experimental configuration that a circularly polarized light beam is impinged with a right angle on a cleaved sidewall of the Fe/x-AlO$_x$/GaAs-based $n$-$i$-$p$ double-heterostructure. The photocurrent $\Delta I$ has showed a well-defined hysteresis loop which resembles that of the magnetization of the in-plane magnetized Fe layer in the devices. The value of $\Delta I$ has been $|\Delta I| \sim 0.2$ nA at 5 K under the remnant magnetization state. Study on temperature dependence of the relative $\Delta I$ value at $H = 0$ has revealed that it is maximized at temperatures 125 – 150 K, and is still measurable at room temperature.

## 1. Introduction

It is expected that circular polarization of light can be quantified electrically with a device incorporating a ferromagnet/semiconductor junction. As shown schematically in Fig. 1(a), spin-polarized carriers, which are generated in the depletion region of a semiconductor $pn$ junction by the illumination with circular polarized (CP) light, yield electromotive force and are transported toward the ferromagnet/semiconductor junction. At the junction, the junction resistance is different between spin-up and spin-down carriers reflecting the different density of states in the ferromagnet[1,2]. This gives rise to a tunnel photocurrent whose magnitude depends on spin polarization of carriers accumulated at the semiconductor side of the junction.

Up to now, relation between CP light and a photocurrent has been studied with various ferromagnet/insulator/semiconductor junctions[2-6]. In those experiments, a CP light beam impinged vertically on the top surface of devices, yielding spin polarized photocarriers whose spin axes are either parallel or anti-parallel to the axis normal to the surface. In order to utilize the spin dependent transport at the ferromagnet/semiconductor junction, ones had to apply vertically a relatively high external magnetic field on a ferromagnetic layer of in-plane anisotropy to force the magnetization vector as well as the spin axis of carriers in the ferromagnet parallel to the spin axis of photo-generated carries.

In this paper, we are concerned with the detection of optical helicity with lateral configuration in which a CP light beam is impinged directly with a right angle on a *cleaved sidewall* of a semiconductor $pn$ junction. In this case, spin axes of photo-generated carriers are always parallel or anti-parallel to the remnant magnetization vector of a ferromagnetic layer of in-plane anisotropy. This fact makes it possible to utilize the spin-dependent

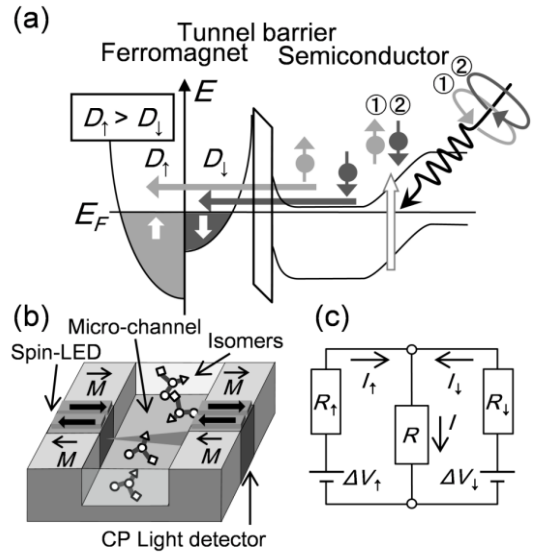

**Fig. 1** (a) Schematic illustration which explains the mechanism of helicity dependent photocurrent. Spin-polarized electrons which are photo-generated in the depletion region of $n$-$i$-$p$ structure (the most right) are transferred towards the ferromagnet/$n$-semiconductor interface, and tunnel into a ferromagnet layer through an ultra-thin insulator barrier. Reflecting the difference in the density of sate between spin-up and spin-down electrons in the ferromagnet, the helicity dependent photocurrent is expected. (b) Schematic illustration of chiral resolution in a micro-channel using a circularly polarized light source (spin-LED) and a circular polarized light detector. Dual spin-injection and spin-detection electrodes are placed, respectively, on spin-LED and CP light detector to determine quantitatively the magnitude of circular polarization. (c) the equivalent circuit representing the mechanism of helicity-dependent photocurrent.

transport channel without the application of an external magnetic field on the device. Another advantage of the lateral configuration is the feasibility of integration into the planer-type optoelectronic circuits. We report in this paper the presence of helicity-dependent photocurrent for wide range of temperature including room temperature (RT) in the experiment in which CP light beam has been impinged on the cleaved sidewall of the ferromagnet/insulator/ semiconductor structure consisting of Fe/x-AlO$_x$/$n$-AlGaAs/ $i$-InGaAs/$p$-AlGaAs/$p$-GaAs(001).

As exemplified by the chiral resolution in synthetic chemistry[7], circularly polarized ellipsometry-based tomography[8,9], and quantum optical computing and information processing[3,10,11], CP photons play an important role in some of the optic and photonic applications. It is expected that a monolithic-type detector of CP photons, together with a monolithic-type CP light emitters[12], would simplify and miniaturize the detection and generation apparatus for circular polarization, and would lead us to the study of new applications: one of such examples is the in-situ detection of optical isomers in a micro-channel, as shown schematically in Fig. 1(b).

## 2. A phenomenological model

The mechanism of helicity-dependent photocurrent can be illustrated schematically by the equivalent circuit shown in Fig. 1(c). The circuit consists of two photo-batteries which generate electromotive force $\Delta V_\uparrow$ and $\Delta V_\downarrow$ per a unit power of incident light for right and left circularly polarized light, respectively, spin-dependent resistance of a tunnel junction incorporating a ferromagnetic layer $R_\uparrow$ and $R_\downarrow$ for which $R_\uparrow \neq R_\downarrow$, and a shunt resistance $R$. The photocurrent $I$ sent to the shunt resistance is expressed as $I = I_\uparrow + I_\downarrow = \Delta V_\uparrow / (R_\uparrow + R) + \Delta V_\downarrow / (R_\downarrow + R)$. When the incident light is linearly polarized, $\Delta V_\uparrow = \Delta V_\downarrow \equiv \Delta V_0$ and $I_0 = \{1 / (R_\uparrow + R) + 1 / (R_\downarrow + R)\}\cdot\Delta V_0$. In the case of CP light, $\Delta V_\uparrow \neq \Delta V_\downarrow$, and we obtain $I_{circ} = \Delta V_\uparrow / (R_\uparrow + R) + \Delta V_\downarrow / (R_\downarrow + R) = (\Delta V_0 + \delta V) / (R_\uparrow + R) + (\Delta V_0 - \delta V) / (R_\downarrow + R)$ in which $\pm\delta V$ is the amount of change in the electromotive force due to circular polarization. Microscopically, $2\delta V$ is the difference in quasi-Fermi level between spin-up and spin-down carriers accumulated at the semiconductor side of the tunnel junction. Consequently, the helicity-dependent component $\Delta I = I_{circ} - I_0$ is expressed as $\Delta I = \{(\Delta V_0 + \delta V) / (R_\uparrow + R) + (\Delta V_0 - \delta V) / (R_\downarrow + R)\} - \{1 / (R_\uparrow + R) + 1 / (R_\downarrow + R)\}\cdot\Delta V_0 = \{1 / (R_\uparrow + R) - 1 / (R_\downarrow + R)\}\cdot\delta V$. The $\delta V$ value is approximately proportional to the spin polarization of carriers $P_{spin}$ which is connected with circular polarization of incident light by the equation $P_{spin} = \gamma\cdot 0.5\cdot P_{circ}$. Here, the correction factor $\gamma$ comes from the degradation of spin polarization during the spin transport from the semiconductor $pn$ junction to the tunnel junction, whereas the proportional constant 0.5 comes from the difference in the spin-dependent optical transition probability between two transitions, the one

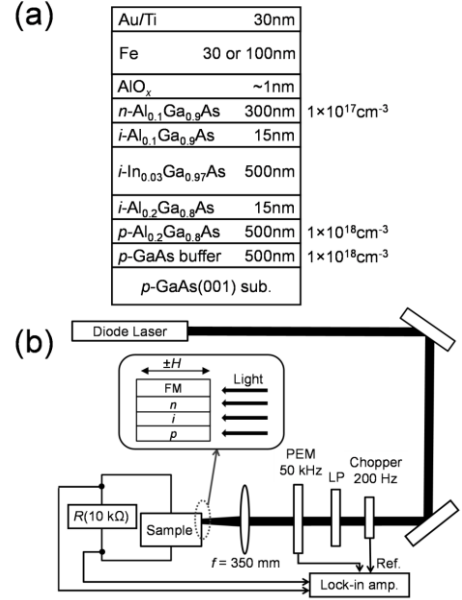

**Fig. 2** (a) Schematic device structure of a cleaved surface. (b) Schematic illustration of experimental setups. Note that circularly polarized light beam has been impinged with a right angle on the cleaved sidewall. We call this setup as the lateral-light-illumination configuration in the text.

from the heavy-hole band state to the conduction band and the other from the light-hole band state to the conduction band state in the case of a III-V compound semiconductor [12,13]. For the experiment using a fully circular-polarized light beam ($P_{circ} = 1$), the helicity-dependent photocurrent $\Delta I$ is expressed by the eq. (1):

$$\Delta I = \left(\frac{1}{R_\uparrow + R} - \frac{1}{R_\downarrow + R}\right)\cdot \delta V \approx \left(\frac{1}{R_\uparrow + R} - \frac{1}{R_\downarrow + R}\right)\cdot (C\cdot P_{spin})$$

$$= \left(\frac{1}{R_\uparrow + R} - \frac{1}{R_\downarrow + R}\right)\cdot (C\cdot \gamma \cdot 0.5 \cdot P_{circ}) \quad (1)$$

Rigorously stated, $\delta V$ should be formulated on the basis of carrier and spin transport with local equilibrium picture[14]. In the present work, we simply take phenomenological approach by connecting $\delta V$ and $P_{spin}$ with a proportional constant $C$.

## 3. Experimental

Shown in Fig. 2(a) is a schematic diagram of a cleaved (1̄10) surface which consists of three different regions: (i) a AlGaAs/InGaAs double-heterostructure (DH) for spin generation and transport, (ii) an ultra-thin, crystalline AlO$_x$ layer to circumvent the conduction mismatch[15], and (iii) a ferromagnetic Fe layer for spin detection. The AlO$_x$/DH in the sample was grown on a $p$-type GaAs (001) substrate by molecular beam epitaxy. The DH consisting of 300-nm $n$-Al$_{0.1}$Ga$_{0.9}$As (Sn ~ 1×10$^{17}$ cm$^{-3}$) / 15-nm undoped Al$_{0.1}$Ga$_{0.9}$As / 500-nm undoped

In$_{0.03}$Ga$_{0.97}$As / 20-nm undoped Al$_{0.2}$Ga$_{0.8}$As / 500-nm p-Al$_{0.2}$Ga$_{0.8}$As (Be ~ 1×10$^{18}$ cm$^{-3}$) / 500-nm p-GaAs (Be ~ 1×10$^{18}$ cm$^{-3}$) / p-GaAs (001), whereas the crystalline AlO$_x$ tunnel barrier (x-AlO$_x$) was formed at RT[15]. After the formation of the crystalline AlO$_x$ layer, the AlO$_x$/DH sample was taken out into an air atmosphere, and transferred to a separate electron beam evaporation system in which both Fe and 30-nm Au/Ti protection layers were deposited. Samples were then post-annealed at 230 °C for 1 hour under a nitrogen atmosphere. Samples having the Fe layer thickness of 30 nm (device A) and 100 nm (device B) were tested in this work. This difference in the Fe layer thickness has resulted in the difference in the coercive field[16].

Samples were then cleaved into 1 mm × 2 mm chips, and placed on a Cu chip carrier. Cu micro-probes, which were pre-assembled in the chip carrier, were pressed on the surfaces of the tested chips to establish electrical contacts. No silver paste was used. The chip carrier was then loaded into a GM-refrigeration optical cryostat with the cleaved (1̄10) sidewall of the chips facing a cryostat window.

Experimental setup is shown schematically in Fig. 2(b). Each chip was connected with the shunt resistance of 10 kΩ across which the voltage due to the zero-bias photo-generated current was measured. This $R$ value was chosen empirically in order to avoid the situation $R \gg R_\uparrow, R_\downarrow$ or $R \ll R_\uparrow, R_\downarrow$. According to eq. (1), measuring the dependence of $R$ on $\Delta I$ would lead us to the estimation of $R_\uparrow$, $R_\downarrow$, and $\delta V$, which we leave for the future study. A linearly polarized cw-laser light with the wavelength $\lambda$ = 785 nm ($h\nu$ = 1.58 eV) and the power $P$ = 20 mW was modulated by the photo-elastic modulator (PEM) to yield a CP light beam of alternating helicity, and was impinged with a right angle on the cleaved (1̄10) sidewall of a chip. The spot size of the focused CP light beam was 1 mm in diameter. Note that the effect due to magnetic circular dichroism of the spin detection layer was completely omitted because the CP light beam did not pass through the Fe layer. Helicity-dependent photocurrent $\Delta I$, which has been defined as the difference between the photocurrent obtained with right- and left-CP light beams, was measured by the double lock-in technique referring the optical chopper of 200 Hz and PEM modulation frequency of 50 kHz[17, 18]. An external magnetic field $H$ was applied along the <1̄10> axis when it was necessary to do so.

## 4. Results and discussion

Helicity-dependent photocurrent component $\Delta I$ has been clearly observed in the light-lateral-illumination configuration. Shown in Figs. 3(a) and (b) are the magnetic field dependence of $\Delta I$ of device A and device B at 5 K, respectively, together with the 5-K magnetization hysteresis curves measured by a magnetometer equipped with the superconducting quantum interference device (SQUID). The polarity of $\Delta I$ switches between positive and negative with sweeping external magnetic fields, which resembles magnetization hysteresis loops of constituent Fe layers. This fact strongly suggests that the $\Delta I$ is attributed to spin-dependent transport across the ferromagnet/insulator/semiconductor junction. The observed discrepancy in the switching field between $\Delta I$ and magnetization, particularly for the device A, can presumably be understood in terms of the difference in a sampling area between the two measurements: magnetization measurement detects the overall magnetization throughout the entire chip area, whereas optical measurement only detects magnetization near the edge of the device. In other words, CP photons are absorbed within the depth of a few μm from the cleaved surface in which spin-polarized carriers are generated and accumulated at the Fe/x-AlO$_x$/n-AlGaAs junction. The influence of magnetic pinning is inferred to be stronger in the region near the cleaved edge than in the region far from the edge, which gives rise to the discrepancy in the switching field. The influence of local magnetic pinning appears to be suppressed in the device B in which the Fe layer is relative thick. It is also worth emphasizing that noticeable $\Delta I$ values has been obtained under the remnant magnetization state: $|\Delta I| \sim 0.2$ nA which corresponds to the electromotive force of around 2 μV at 5 K. This is very encouraging in that the application of an external magnetic field can be avoided in the light-lateral-illumination configuration.

Figure 4 (a) shows the temperature dependence of $\Delta I/I - H$ curves for device B. Here, $I$ represents the overall photocurrent incorporating the non-helicity-dependent component. To our surprise, hysteretic $\Delta I/I - H$ curves have been observed up to RT,

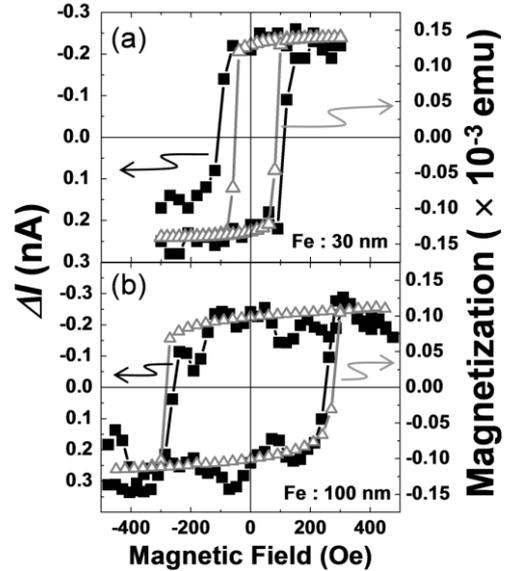

**Fig. 3** Field dependence of helicity-dependent photo-current component $\Delta I$ and magnetization for (a) device A and (b) device B.

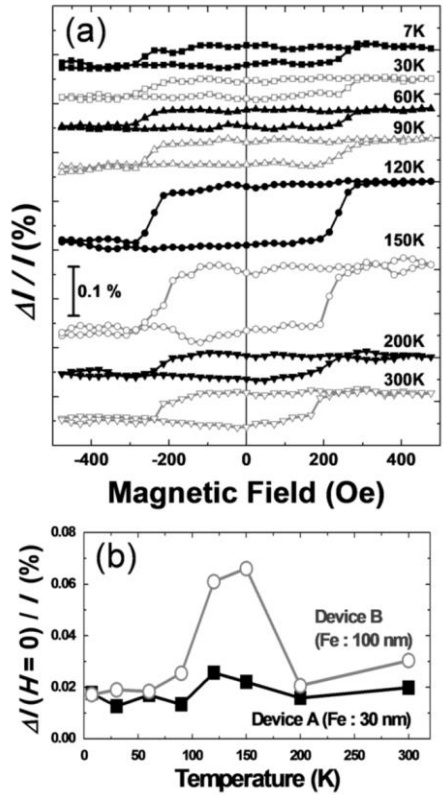

**Fig. 4** (a) $\Delta I/I - H$ hysteresis loops obtained at different temperatures for device B. (b) Temperature dependence of $\Delta I/I$ at $H = 0$ for device A and device B.

suggesting that spin-polarized photo-generated carriers have survived in the DH and have reached the Fe/x-AlO$_x$ junction even at RT. The magnitude of $\Delta I/I$ at $H = 0$ is summarized in Fig. 4(b) for device A and device B both of which qualitatively exhibit similar behavior. The $\Delta I/I$ values are relative small at low temperatures (< 90 K), but increase with increasing temperature above around 100 K and show maximum at 125-150 K, beyond which they turn to the reduction. We infer that the observed complicated behavior, being difficult to simply explain by the established spin relaxation process in the bulk $n$-GaAs[19], could be explained by taking into account of the temperature dependences of band gap energy and relative band alignment around the Fermi level, as well as D'yakonov-Perel' spin relaxation. The development of quantitative model is needed to further elucidate the observed experimental results and to find the way to enhance the electrical signal due to helicity-dependent component.

## 5. Conclusions

Helicity-dependent photocurrent $\Delta I$ has been observed successfully in the light-lateral-illumination configuration in which a CP light beam is impinged with a right angle on a cleaved sidewall of the Fe/x-AlO$_x$/GaAs-based $n$-$i$-$p$ double-heterostructure grown on a $p$-GaAs(001) substrate. The helicity-dependent photocurrent $\Delta I$ exhibits a well-defined hysteretic characteristic which resembles the magnetization hysteresis loop of the in-plane magnetized Fe layer in the tested devices. Study on temperature dependence of relative $\Delta I$ value ($\Delta I/I$) at $H = 0$ has revealed that $\Delta I/I$ is maximized at temperatures 125 – 150 K, and is still detectable at RT.

**Acknowledgements** We acknowledge partial supports from Advanced Photon Science Alliance Project from MEXT and Grant-in-Aid for Scientific Research (No. 22226002) from JSPS.